# Analysis of the quantum-mechanical equivalence of the metrics of a centrally symmetric uncharged gravitational field

M.V. Gorbatenko, V.P. Neznamov[1]

RFNC-VNIIEF, 37 Mira Ave., Sarov, 607188, Russia

Abstract

In the paper we analyze the quantum-mechanical equivalence of the metrics of a centrally symmetric uncharged gravitational field. We consider the Schwarzschild metrics in the spherical, isotropic and harmonic coordinates, and the Eddington-Finkelstein, Painleve-Gullstrand, Lemaitre-Finkelstein, Kruskal metrics. The scope of the analysis includes domains of the wave functions of Dirac's equation, hermiticity of Hamiltonians, and the possibility of existence of stationary bound states of spin-half particles.

The constraint on the domain of the wave functions of the Hamiltonian in a Schwarzschild field in spherical coordinates $(r > r_0)$ resulting from the fulfillment of Hilbert's condition $g_{00} > 0$ also holds in other coordinates for all the metrics considered.

The self-adjoint Hamiltonians for the Schwarzschild metrics in the spherical, isotropic and harmonic coordinates and also for the Eddington-Finkelstein and Painleve-Gullstrand metrics are Hermitian, and for them the existence of stationary bound states of spin-half particles is possible.

The self-adjoint Hamiltonians for non-stationary Lemaitre-Finkelstein and Kruskal metrics have the explicit dependence on the temporal coordinates and stationary bound states of spin-half particles cannot be defined for these Hamiltonians.

The results of this study can be useful when addressing the issues related to the evolution of the universe and interaction of collapsars with surrounding matter.

---

[1] E-mail: neznamov@vniief.ru

## 1. Introduction

The Schwarzschild metric [1] is a widely known solution of general relativity for a point uncharged centrally symmetric gravitational field.

The classical Schwarzschild solution is characterized by a spherically symmetric point source of gravitational field of mass $M$ and an "event horizon" (gravitational radius)

$$r_0 = \frac{2GM}{c^2}. \tag{1}$$

In (1), $G$ is the gravitational constant, and $c$ is the speed of light. In the classical case, as seen by a distant observer, a test particle reaches the "event horizon" in an infinitely long time.

There are a number of other solutions derived by coordinate transformations of the Schwarzschild solution and also representing exact solutions of general relativity.

The following solutions can be mentioned: the Schwarzschild metric in isotropic coordinates [2], the Schwarzschild metric in harmonic coordinates [3], the Lemaitre-Finkelstein metric [4], [5], the Kruskal metric [6], [7], the Eddington-Finkelstein metric [8], [5], and the Painleve-Gullstrand metric [9], [10].

In [11] - [13], we developed a method for deriving self-adjoint Dirac Hamiltonians with a flat scalar product of the wave functions within the framework of pseudo-Hermitian quantum mechanics for arbitrary, including time dependent, gravitational fields.

It follows from single-particle quantum mechanics that if a Hamiltonian is Hermitian with corresponding equality of scalar products of the wave functions $\left((\Phi, H\Psi) = (H\Phi, \Psi)\right)$ and if boundary conditions are established, time-independent self-adjoint Hamiltonians $\left(H = H^+\right)$ should provide for the existence of stationary bound states of spin-half particles with a real energy spectrum[2].

In this study, we explore the quantum-mechanical equivalence of the above centrally symmetric solutions of the general relativity equations obtained by coordinate transformations of the Schwarzschild metric [1]. For each metric, we analyze Dirac's self-adjoint Hamiltonians with a flat scalar product of the wave functions. We examine domains of the wave functions, hermiticity of Hamiltonians, and the possibility of existence of stationary bound states of spin-half particles.

---

[2] Not every self-adjoint Hamiltonian will be Hermitian for the definitions adopted. For a Hamiltonian to be Hermitian, the wave functions should behave correspondingly to ensure the equality $\left((\Phi, H\Psi) = (H\Phi, \Psi)\right)$.

For each metric, Hamiltonians are obtained both directly with tetrads in the Schwinger gauge [14], and through coordinate transformations and the Lorenz transformations of the self-adjoint Hamiltonian in a Schwarzschild gravitational field [1].

Section 2 presents the methodology of analysis of the quantum-mechanical equivalence of the metrics of a centrally symmetric uncharged gravitational field.

In Sections 3 – 5, we analyze the self-adjoint Hamiltonians in the Schwarzschild fields with isotropic and harmonic coordinates, in the Eddington-Finkelstein and Painleve-Gullstrand fields and in the Lemaitre-Finkelstein and Kruskal fields.

In the Conclusion we discuss the results of the quantum-mechanical analysis.

**2. Methodology of analysis of the quantum-mechanical equivalence of the centrally symmetric solutions of the general relativity equations**

**2.1. Dirac's equation**

It is assumed that motion of a spin-half particle in an external gravitational field is described by covariant Dirac's equation. In the units of $\hbar = c = 1$, it is written in the form

$$\gamma^\alpha \nabla_\alpha \psi - m\psi = 0. \tag{2}$$

Here, $m$ is the particle mass, $\psi$ is the four-component bispinor, $\nabla_\alpha$ is the covariant derivative, and $\gamma^\alpha$ are global 4x4 Dirac's matrices satisfying the relationship

$$\gamma^\alpha \gamma^\beta + \gamma^\beta \gamma^\alpha = 2g^{\alpha\beta} E. \tag{3}$$

In (3), $g^{\alpha\beta}$ is the inverse metric tensor; $E$ is a 4x4 identity matrix.

In (2), (3) and below, the Greek symbols assume the values of (0, 1, 2, 3), and the symbols from the Roman alphabet assume the values of (1, 2, 3). The corresponding terms with the same superscripts and subscripts are understood to be summed up.

Below, in addition to Dirac's matrices $\gamma^\alpha$ with global indices, we will use Dirac's matrices $\gamma^{\underline{\alpha}}$ with local indices satisfying the relationship

$$\gamma^{\underline{\alpha}} \gamma^{\underline{\beta}} + \gamma^{\underline{\beta}} \gamma^{\underline{\alpha}} = 2\eta^{\underline{\alpha}\underline{\beta}} E. \tag{4}$$

In (4), $\eta^{\underline{\alpha}\underline{\beta}}$ corresponds to the metric tensor of flat Minkowski space with a signature

$$\eta_{\underline{\alpha}\underline{\beta}} = diag[1, -1, -1, -1]. \tag{5}$$

It is convenient to choose the quantities $\gamma^{\underline{\alpha}}$ such that they have the same form for all local frames of reference. Both systems $\gamma^{\underline{\alpha}}$ and $\gamma^{\alpha}$ can be used to construct a full system of 4x4 matrices. An example of a full system is given below:

$$E, \gamma^{\alpha}, S^{\alpha\beta} = \frac{1}{2}\left(\gamma^{\alpha}\gamma^{\beta} - \gamma^{\beta}\gamma^{\alpha}\right),$$
$$\gamma_5 = \gamma^0 \gamma^1 \gamma^2 \gamma^3, \quad \gamma_5 \gamma^{\alpha}. \tag{6}$$

Any set of Dirac's matrices is suitable for several discrete automorphisms. We restrict ourselves to the automorphism

$$\gamma^{\alpha} \to \left(\gamma^{\alpha}\right)^{+} = -D\gamma^{\alpha} D^{-1}, \tag{7}$$

the matrix $D$ is called anti-Hermitizing.

The covariant derivative of the bispinor $\nabla_{\alpha}\psi$ in (2) equals

$$\nabla_{\alpha}\psi = \frac{\partial \psi}{\partial x^{\alpha}} + \Phi_{\alpha}\psi. \tag{8}$$

In (8), to define the bispinor connectivities $\Phi_{\alpha}$, we should choose a certain system of tetrad vectors $H_{\underline{\alpha}}^{\mu}$ satisfying the relationships

$$H_{\underline{\alpha}}^{\mu} H_{\underline{\beta}}^{\nu} g_{\mu\nu} = \eta_{\underline{\alpha}\underline{\beta}}. \tag{9}$$

In addition to the tetrad vectors $H_{\underline{\alpha}}^{\mu}$, one can introduce three other systems of tetrad vectors $H_{\underline{\alpha}\mu}, H^{\underline{\alpha}\mu}, H_{\mu}^{\underline{\alpha}}$, which differ from $H_{\underline{\alpha}}^{\mu}$ in the position of the global and local (underlined) indices. The global indices are raised up and lowered by means of the metric tensor $g_{\mu\nu}$ and inverse tensor $g^{\mu\nu}$, and the local ones, by means of the tensors $\eta_{\underline{\alpha}\underline{\beta}}, \eta^{\underline{\alpha}\underline{\beta}}$.

When choosing the system of tetrad vectors, the bispinor connectivities are defined by means of Christoffel derivatives of the tetrad vectors

$$\Phi_{\alpha} = -\frac{1}{4} H_{\mu}^{\underline{\varepsilon}} H_{\nu\underline{\varepsilon};\alpha} S^{\mu\nu} = \frac{1}{4} H_{\underline{\mu}}^{\varepsilon} H_{\underline{\nu}\varepsilon;\alpha} S^{\underline{\mu}\underline{\nu}}. \tag{10}$$

The relationship between $\gamma^{\alpha}$ and $\gamma^{\underline{\alpha}}$ is given by the expression

$$\gamma^{\alpha} = H_{\underline{\beta}}^{\alpha} \gamma^{\underline{\beta}}. \tag{11}$$

At coordinate transformations

$$\{x^{\alpha}\} \to \{x'^{\alpha}\} \tag{12}$$

the following relationships hold:

$$\gamma'^{\alpha} = \frac{\partial x'^{\alpha}}{\partial x^{\beta}} \gamma^{\beta}, \tag{13}$$

$$\Phi'_\alpha = \frac{\partial x'^\beta}{\partial x^\alpha} \Phi_\beta. \tag{14}$$

Two arbitrary systems of tetrad vectors in the same space-time are related to each other by the Lorentz transformation $L(x)$

$$\tilde{H}^\mu_{\underline{\alpha}}(x) = \Lambda^{\underline{\beta}}_{\underline{\alpha}}(x) H^\mu_{\underline{\beta}}(x). \tag{15}$$

The quantities $\Lambda^{\underline{\beta}}_{\underline{\alpha}}(x)$ satisfy the relationships

$$\Lambda^{\underline{\mu}}_{\underline{\alpha}}(x)\Lambda^{\underline{\nu}}_{\underline{\beta}}(x)\eta^{\underline{\alpha\beta}} = \eta^{\underline{\mu\nu}},$$
$$\Lambda^{\underline{\mu}}_{\underline{\alpha}}(x)\Lambda^{\underline{\nu}}_{\underline{\beta}}(x)\eta_{\underline{\mu\nu}} = \eta_{\underline{\alpha\beta}}. \tag{16}$$

The mathematical apparatus introduced above ensures the covariance of Dirac's equation (2) both at the coordinate transformations (12), and with the transition from one system of tetrad vectors to another (15).

### 2.2. Schrödinger relativistic equation

To completely utilize the quantum mechanics apparatus, it is reasonable to move from Dirac's equation (2) to a Schrödinger-type equation with separation of the time derivative of the wave function.

$$i\frac{\partial \psi}{\partial t} = H\psi. \tag{17}$$

On the left side of (17), $t = x^0$; on the right side of (17), $H$ is the Hamiltonian operator.

Considering (8) and the equality $\gamma^0\gamma^0 = g^{00}$, one can obtain an expression for the Hamiltonian from (2):

$$H = \frac{m}{g^{00}}\gamma^0 - \frac{1}{g^{00}}i\gamma^0\gamma^k\frac{\partial}{\partial x^k} - i\Phi_0 - \frac{1}{g^{00}}i\gamma^0\gamma^k\Phi_k. \tag{18}$$

In [12] we show that, in the same space-time, one can move from any system of tetrad vectors $\{H^\mu_{\underline{\alpha}}(x)\}$ to a system of tetrad vectors $\{\tilde{H}^\mu_{\underline{\alpha}}(x)\}$ in the Schwinger gauge [14] by the Lorentz transformation $L(x)$.

For the system $\{\tilde{H}^\mu_{\underline{\alpha}}(x)\}$,

$$\tilde{H}^0_{\underline{0}} = \sqrt{g^{00}}; \quad \tilde{H}^k_{\underline{0}} = -\frac{g^{0k}}{\sqrt{g^{00}}}; \quad \tilde{H}^0_{\underline{k}} = 0. \tag{19}$$

Spatial tetrads satisfying the relationships written below can be used as tetrad vectors $\tilde{H}^n_{\underline{m}}$:

$$\tilde{H}^m_{\underline{k}}\tilde{H}^n_{\underline{k}} = f^{mn}; \quad f^{mn} = g^{mn} + \frac{g^{om}g^{0n}}{g^{00}}; \quad f^{mn}g_{nk} = \delta^m_k. \tag{20}$$

The Lorentz transformation matrix is written as

$$L(x) = R \cdot \exp\left\{\frac{\theta}{2}\frac{\tilde{H}^\mu_{\underline{0}}H^\nu_{\underline{0}}S_{\mu\nu}}{\sqrt{\left(\tilde{H}^\varepsilon_{\underline{0}}H_{\underline{0}\varepsilon}\right)^2 - 1}}\right\}. \tag{21}$$

Here, $R$ is the spatial rotation matrix commuting with $\gamma^{\underline{0}}$. The second multiplier is transformation of hyperbolic rotation (boost) to an angle $\theta$ determined from the relationship

$$\text{th}\frac{\theta}{2} = \sqrt{\frac{\left(\tilde{H}^\varepsilon_{\underline{0}}H_{\underline{0}\varepsilon}\right)+1}{\left(\tilde{H}^\varepsilon_{\underline{0}}H_{\underline{0}\varepsilon}\right)-1}}. \tag{22}$$

The matrix $L$ transforms $\gamma^0(x)$ into the following form

$$L\gamma^0 L^{-1} = \sqrt{g^{00}}\gamma^{\underline{0}}. \tag{23}$$

Considering some freedom of choice of spatial tetrads $\tilde{H}^n_{\underline{m}}$ determined by the relationships (20), in moving from the Hamiltonian (18) with tetrad vectors $\{H^\mu_{\underline{\alpha}}(x)\}$ to the Hamiltonians $\tilde{H}$ with a system of tetrad vectors in the Schwinger gauge $\{\tilde{H}^\mu_{\underline{\alpha}}(x)\}$ and with various sets of $\tilde{H}^n_{\underline{m}}$, one can obtain distinct expressions. As a matter of fact, these Hamiltonians are physically equivalent, because they are related by unitary matrices of spatial rotations.

### 2.3. Hermiticity conditions for Hamiltonians and wave functions

In [12] we show that stationary Dirac's Hamiltonians in an external gravitational field are pseudo-Hermitian and satisfy the condition of pseudo-Hermitian quantum mechanics [15] - [17].

$$H^+ = \rho H \rho^{-1}. \tag{24}$$

The operator $\rho$ in (24) is Parker's weight operator [18]

$$\rho = \sqrt{-g}\gamma^0\gamma^{\underline{0}}, \tag{25}$$

where $g$ is the determinant of metric $g_{\mu\nu}$.

For tetrad vectors in the Schwinger gauge

$$\rho = \sqrt{-g}\sqrt{g^{00}}. \tag{26}$$

The scalar product of the wave functions with the operator $\rho$ takes the form

$$(\Phi, \Psi) = \int \psi^+(x)\rho(x)\psi(x)d^3x. \tag{27}$$

The general condition of hermiticity for Dirac's Hamiltonians in external gravitational fields $(\Phi, H\Psi) = (H\Phi, \Psi)$ can be written in the form [11]

$$\oint ds_k \left( \sqrt{-g}\, j^k \right) + \int d^3 x \sqrt{-g} \left[ \psi^+ \gamma^{\underline{0}} \left( \gamma^0_{,0} + \begin{pmatrix} 0 & 0 \\ 0 & 0 \end{pmatrix} \gamma^0 \right) \psi + \begin{pmatrix} 0 & k \\ k & 0 \end{pmatrix} j^0 \right] = 0. \tag{28}$$

In (28), current components $j^\mu$ are defined as

$$j^\mu = \psi^+ \gamma^{\underline{0}} \gamma^\mu \psi. \tag{29}$$

For the time-independent Hamiltonians $\gamma^0_{,0} \equiv \dfrac{\partial \gamma^0}{\partial x^0} = 0$, the Christoffel symbols $\begin{pmatrix} 0 \\ 0\ 0 \end{pmatrix}, \begin{pmatrix} k \\ k\ 0 \end{pmatrix}$ for centrally symmetric fields are zero, and the condition (28) becomes equal to

$$\oint ds_k \left( \sqrt{-g}\, j^k \right) = 0. \tag{30}$$

If there exists an operator $\eta$ satisfying the relationship

$$\left( \frac{g_G}{g} \right)^{1/2} \rho = \eta^+ \eta, \tag{31}$$

then the Hamiltonian

$$H_\eta = \eta H \eta^{-1} \tag{32}$$

will be self-adjoint,

$$H_\eta^+ = H_\eta, \tag{33}$$

and the scalar product (27) becomes flat (without weight factor $\rho(x)$).

In this case,

$$\psi_\eta(x) = \eta \psi(x). \tag{34}$$

In (31), we introduce notation $g_G = \dfrac{g}{g_c}$ [13], in which $g_c$ is a determinant that occurs when the volume element is written in curvilinear coordinates ($g_c = 1$ for the Cartesian coordinates, $g_c = r^2$ for the cylindrical coordinates, $g_c = r^4 \sin^2 \theta$ for the spherical coordinates, etc.).

In [13] we show that, for the centrally symmetric gravitational field metrics of interest, the Hamiltonian $H_\eta$ (32) can be obtained without direct calculation of the bispinor connectivities (10) from the expression

$$H_\eta = \frac{1}{2} \left( \tilde{H}_{red} + \tilde{H}_{red}^+ \right), \tag{35}$$

where $\tilde{H}_{red}$ is part of the initial Hamiltonian (18) with tetrads in the Schwinger gauge without summands with bispinor connectivities $\tilde{\Phi}_0, \tilde{\Phi}_k$.

## 2.4. Domains of wave functions

When defining the domains of the wave functions, we will be guided by the fulfillment of Hilbert's causality conditions [19], [20]

$$g < 0;\ g_{00} > 0;\ g_{11} < 0;\ \begin{vmatrix} g_{11} & g_{12} \\ g_{21} & g_{22} \end{vmatrix} > 0;\ \begin{vmatrix} g_{11} & g_{12} & g_{13} \\ g_{21} & g_{22} & g_{23} \\ g_{31} & g_{32} & g_{33} \end{vmatrix} < 0. \tag{36}$$

Special attention will be paid to the fulfillment of the second inequality $g_{00} > 0$; all the other inequalities in (36) are generally satisfied for the known general relativity solutions.

## 2.5 Inertial and rotating frames of reference in Minkowski space

As an illustration of the necessity of satisfying the condition $g_{00} > 0$, we consider Dirac's Hamiltonians in the inertial and the rotating frames of reference of Minkowski space.

For the inertial frame of reference $(x'^\mu) = (t', x', y', z')$, the Hamiltonian of a Dirac particle with an unrestricted domain of the wave functions is given by

$$H' = \gamma^0 m - i\gamma^0 \gamma^k \frac{\partial}{\partial x'^k}. \tag{37}$$

We introduce a rotating frame of reference [20],

$$t = t';\ x = x'\cos\omega t + y'\sin\omega t;\ y = -x'\sin\omega t + y'\cos\omega t;\ z = z'. \tag{38}$$

In (38), the velocity of rotation $\omega$ is a real number. The Minkowski metric in this frame of reference is stationary and is written in the form

$$ds^2 = \left(1 - \omega^2(x^2 + y^2)\right)dt^2 + 2\omega(ydx - xdy)dt - dx^2 - dy^2 - dz^2. \tag{39}$$

The Hamiltonian (37) in the new frame of reference takes the form (see, e.g., [21])

$$H = \gamma^0 m - i\gamma^0 \gamma^k \frac{\partial}{\partial x'^k} - i\omega\left(y\frac{\partial}{\partial x} - x\frac{\partial}{\partial y}\right). \tag{40}$$

The domain of the wave functions of the Hamiltonian (40) is constrained by the condition $g_{00} > 0$, which, for the metric (39), resolves into the condition $\sqrt{x^2 + y^2} < \frac{1}{\omega}$. If this condition were not satisfied, the velocity of rotation for distances $\sqrt{x^2+y^2}>\frac{1}{\omega}$ would exceed the speed of light. Thus, at $g_{00}<0$, the rotational frame of reference cannot be the case for real bodies [20].

**2.6. Roadmap of quantum-mechanical analysis of the equivalence of the centrally symmetric solutions of the general relativity equations**

As a basic metric we consider the Schwarzschild solution in the coordinates $(t,r,\theta,\varphi)$. All the other centrally symmetric solutions of the general relativity equations will be obtained by corresponding coordinate transformations of the basic metric.

For each metric, we will directly obtain self-adjoint Hamiltonians with a flat scalar product of the wave functions and tetrads in the Schwinger gauge (19), (20).

In addition for the transformed matrices, we will further obtain self-adjoint Hamiltonians in the $\eta$-representation and with the tetrads (19), (20) in two steps. Step 1 includes transformation of the basic self-adjoint Schwarzschild Hamiltonian to the coordinates of transformed metric in accordance with (11) – (13) while preserving the tetrads of the basic Hamiltonian, i.e. the tetrads with such transformation will equal

$$H'^{\alpha}_{\underline{\beta}} = \frac{\partial x'^{\alpha}}{\partial x^{\mu}} H^{\mu}_{\underline{\beta}}. \tag{41}$$

If necessary, the second step is performed, which includes Lorentz transformation (15), (16), (21), (22) to bring the resulting Hamiltonian in the coordinates of transformed metric to the tetrads in the Schwinger gauge.

At the end of the transformations, we check the domain of the wave functions, hermiticity of the Hamiltonian, and the possibility of existence of stationary bound states of spin-half particles in the corresponding gravitational fields.

**3. Schwarzschild solution in the $(t,r,\theta,\varphi)$ coordinates**

The square of interval is

$$ds^2 = f_S dt^2 - \frac{dr^2}{f_S} - r^2\left(d\theta^2 + \sin^2\theta d\varphi^2\right). \tag{42}$$

In (42), $f_S = 1 - \frac{r_0}{r}$.

The non-zero tetrads in the Schwinger gauge $\tilde{H}^{\mu}_{\underline{\alpha}}$ equal

$$\tilde{H}^0_{\underline{0}} = \frac{1}{\sqrt{f_S}}; \quad \tilde{H}^1_{\underline{1}} = \sqrt{f_S}; \quad \tilde{H}^2_{\underline{2}} = \frac{1}{r}; \quad \tilde{H}^3_{\underline{3}} = \frac{1}{r\sin\theta}. \tag{43}$$

Following (11), the matrices $\tilde{\gamma}^\alpha$ equal

$$\tilde{\gamma}^0 = \frac{1}{\sqrt{f_S}}\gamma^0; \quad \tilde{\gamma}^1 = \sqrt{f_S}\gamma^1; \quad \tilde{\gamma}^2 = \frac{1}{r}\gamma^2; \quad \tilde{\gamma}^3 = \frac{1}{r\sin\theta}\gamma^3. \tag{44}$$

Dirac's self-adjoint Hamiltonian with the tetrads (43) is written in the form [13]

$$H_\eta = \sqrt{f_S}m\gamma^0 - i\sqrt{f_S}\gamma^0\left\{\gamma^1\sqrt{f_S}\left(\frac{\partial}{\partial r}+\frac{1}{r}\right)+\gamma^2\frac{1}{r}\left(\frac{\partial}{\partial\theta}+\frac{1}{2}\text{ctg}\,\theta\right)+\right.$$
$$\left.+\gamma^3\frac{1}{r\sin\theta}\frac{\partial}{\partial\varphi}\right\} - \frac{i}{2}\frac{\partial f_S}{\partial r}\gamma^0\gamma^1. \tag{45}$$

The domain of the wave functions of Dirac's equation with the Hamiltonian (45) is constrained by Hilbert's causality condition

$$g_{00} > 0 \to f_S = 1 - \frac{r_0}{r} > 0 \to r > r_0. \tag{46}$$

It follows from (46) that

$$\sqrt{f_S} \text{ is a positive real number.} \tag{47}$$

The transformation operator $\eta$ (31) equals

$$\eta = \frac{1}{f_S^{1/4}}. \tag{48}$$

The current components

$$j^\mu = \psi_\eta^+ \left(\eta^{-1}\right)^+ \left(\gamma^0\tilde{\gamma}^\mu\right)\left(\eta^{-1}\right)\psi_\eta \tag{49}$$

equal

$$j^0 = \psi_\eta^+\psi_\eta, \tag{50}$$

$$j^r = \psi_\eta^+ f_S\gamma^1\psi_\eta = 0, \tag{51}$$

$$j^\theta = \psi_\eta^+ \frac{\sqrt{f_S}}{r}\gamma^2\psi_\eta = 0, \tag{52}$$

$$j^\varphi = \psi_\eta^+ \frac{\sqrt{f_S}}{r\sin\theta}\gamma^3\psi_\eta. \tag{53}$$

The equality to zero of the radial (51) and the polar (52) current components is attributed to the form of the spherical harmonics for spin-half (see, e.g., [22], [23]).

In case of the Schwarzschild metric, the hermiticity condition for Dirac's Hamiltonians (28), (30) for the domain of the wave functions (46) can be written in the form

$$4\pi r^2 j^r(r)\big|_{r\to\infty} + 4\pi r^2 j^r(r)\big|_{r\to r_0} = 0, \tag{54}$$

which, considering (51), is fulfilled automatically. This suggests that if we introduce a physically reasonable boundary condition for the wave functions on any spherical surface with $r > r_0$ and select exponentially negative-going solutions as $r \to \infty$, the self-adjoint Hamiltonian (45) will have a stationary real energy spectrum of bound states of spin-half particles [24], [25].

Sometimes, to implement the possibility of particle motion in the Schwarzschild field under the "event horizon", it is suggested that, in the Schwarzschild metric (42), the temporal and the radial coordinates be mutually interchanged [26], [27]. The square of interval then becomes equal to

$$ds^2 = \frac{t}{r_0 - t} dt^2 - \frac{r_0 - t}{t} dr^2 - t^2 \left( d\theta^2 + \sin^2\theta d\varphi^2 \right). \tag{55}$$

In (55), $t \in (0, r_0)$, $r \in (0, \infty)$.

The non-zero components of the tetrad vectors in the Schwinger gauge equal

$$\tilde{H}^0_{\underline{0}} = \sqrt{\frac{r_0 - t}{t}};\quad \tilde{H}^1_{\underline{1}} = \sqrt{\frac{t}{r_0 - t}};\quad \tilde{H}^2_{\underline{2}} = \frac{1}{t};\quad \tilde{H}^3_{\underline{3}} = \frac{1}{t\sin\theta}. \tag{56}$$

The self-adjoint Hamiltonian in the $\eta$-representation equals

$$H_\eta = \sqrt{\frac{t}{r_0 - t}} m\gamma^{\underline{0}} - i\gamma^{\underline{0}}\gamma^{\underline{1}} \frac{t}{r_0 - t}\left(\frac{\partial}{\partial r} + \frac{1}{r}\right) - i\gamma^{\underline{0}}\gamma^{\underline{2}} \sqrt{\frac{t}{r_0 - t}} \frac{1}{t}\left(\frac{\partial}{\partial \theta} + \frac{1}{2}\mathrm{ctg}\,\theta\right) -$$
$$- i\gamma^{\underline{0}}\gamma^{\underline{3}} \sqrt{\frac{t}{r_0 - t}} \frac{1}{t\sin\theta} \frac{\partial}{\partial \varphi}. \tag{57}$$

The domain of the wave functions of the Hamiltonian (57) is constrained by Hilbert's causality condition $g_{00} > 0$, i.e.

$$t < r_0. \tag{58}$$

The Hamiltonian (57) is explicitly time-dependent and physically non-equivalent to the stationary Hamiltonian (45) with the domain of the wave functions $r > r_0$. Hilbert's condition $g_{00} \neq 0$ does not allow "cross-linking" the wave functions at the "event horizon" $r = r_0$.

Further we will consider transformations of the Hamiltonian (45) with the domain of the wave functions $r > r_0$ and with real positive values of $\sqrt{f_S}$.

# 4. Schwarzschild metrics in isotropic and harmonic coordinates

## 4.1 Schwarzschild solution in isotropic coordinates

The coordinates are

$$(t, R, \theta, \varphi). \tag{59}$$

The coordinate transformation is

$$r = R\left(1 + \frac{r_0}{4R}\right)^2; \quad dR = \frac{dr}{\left(1 - \frac{r_0^2}{16R^2}\right)}. \tag{60}$$

The square of interval is

$$ds^2 = V^2(R)dt^2 - W^2(R)\left[dR^2 + R^2\left(d\theta^2 + \sin^2\theta d\varphi^2\right)\right]. \tag{61}$$

Here,

$$V(R) = \frac{1 - \frac{r_0}{4R}}{1 + \frac{r_0}{4R}}, \quad W(R) = \left(1 + \frac{r_0}{4R}\right)^2. \tag{62}$$

The values $(-g)$, $g_G$ and $\eta$ equal

$$-g = V^2 \cdot W^6 \cdot R^4 \sin^2\theta = \left(1 + \frac{r_0}{4R}\right)^6 R^4 \sin^2\theta, \tag{63}$$

$$g_G = \left(1 + \frac{r_0}{4R}\right)^6, \tag{64}$$

$$\eta = (g_G)^{1/4}(g^{00})^{1/4} = \frac{\left(1 + \frac{r_0}{4R}\right)^2}{\left(1 - \frac{r_0}{4R}\right)^{1/2}}. \tag{65}$$

The non-zero components of the tetrad vectors $\tilde{H}_{\underline{\alpha}}^{\mu}$ in the Schwinger gauge equal:

$$\tilde{H}_{\underline{0}}^0 = \frac{1 + \frac{r_0}{4R}}{1 - \frac{r_0}{4R}}; \quad \tilde{H}_{\underline{1}}^1 = \frac{1}{\left(1 + \frac{r_0}{4R}\right)^2}; \quad \tilde{H}_{\underline{2}}^2 = \frac{1}{\left(1 + \frac{r_0}{4R}\right)^2} \frac{1}{R}; \quad \tilde{H}_{\underline{3}}^3 = \frac{1}{\left(1 + \frac{r_0}{4R}\right)^2} \frac{1}{R\sin\theta}. \tag{66}$$

The matrices $\tilde{\gamma}^\alpha$ equal

$$\tilde{\gamma}^{0} = \frac{1+\frac{r_0}{4R}}{1-\frac{r_0}{4R}}\gamma^{\underline{0}};\ \tilde{\gamma}^{1} = \frac{1}{\left(1+\frac{r_0}{4R}\right)^{2}}\gamma^{\underline{1}};\ \tilde{\gamma}^{2} = \frac{1}{R\left(1+\frac{r_0}{4R}\right)^{2}}\gamma^{\underline{2}};\ \tilde{\gamma}^{3} = \frac{1}{R\sin\theta\left(1+\frac{r_0}{4R}\right)^{2}}\gamma^{\underline{3}}. \quad (67)$$

The self-adjoint Hamiltonian in the $\eta$ - representation with the tetrads (66) equals

$$H_\eta = \frac{1-\frac{r_0}{4R}}{1+\frac{r_0}{4R}}m\gamma^{\underline{0}} - i\frac{1-\frac{r_0}{4R}}{\left(1+\frac{r_0}{4R}\right)^{3}}\gamma^{\underline{0}}\left[\gamma^{\underline{1}}\left(\frac{\partial}{\partial R}+\frac{1}{R}\right)\right]+$$

$$+\gamma^{\underline{2}}\frac{1}{R}\left(\frac{\partial}{\partial\theta}+\frac{1}{2}\mathrm{ctg}\theta\right)+\gamma^{\underline{3}}\frac{1}{R\sin\theta}\frac{\partial}{\partial\theta}\right]-\frac{i}{2}\gamma^{\underline{0}}\gamma^{\underline{1}}\frac{\partial}{\partial R}\frac{1-\frac{r_0}{4R}}{\left(1+\frac{r_0}{4R}\right)^{3}}. \quad (68)$$

The current components

$$j^{\mu} = \psi_\eta^{+}\left(\eta^{-1}\right)^{+}\gamma^{\underline{0}}\tilde{\gamma}^{\mu}\left(\eta^{-1}\right)\psi_\eta \quad (69)$$

equal

$$j^{0} = \psi_\eta^{+}\frac{1}{\left(1+\frac{r_0}{4R}\right)^{3}}\psi_\eta, \quad (70)$$

$$j^{r} = j^{\theta} = 0, \quad (71)$$

$$j^{\varphi} = \psi_\eta^{+}\frac{1-\frac{r_0}{4R}}{\left(1+\frac{r_0}{4R}\right)^{6}R\sin\theta}\gamma^{\underline{0}}\gamma^{\underline{3}}\psi_\eta. \quad (72)$$

Now we obtain the Hamiltonian (68) by direct transformation of the basic Hamiltonian (45) with the tetrads (43).

As a result of the coordinate transformation (60), the tetrads (43) transform in accordance with (41).

$$\left(H_{\underline{0}}^{\prime 0}\right)_{is} = \frac{\partial t}{\partial t}\left(H_{\underline{0}}^{0}\right)_{S} = \frac{1}{\sqrt{f_s}} = \frac{1+\frac{r_0}{4R}}{1-\frac{r_0}{4R}}; \quad (73)$$

$$\left(H_{\underline{1}}^{\prime 1}\right)_{is} = \frac{\partial R}{\partial r}\left(H_{\underline{1}}^{1}\right)_{S} = \frac{\sqrt{f_s}}{1-\frac{r_0^2}{16R^2}} = \frac{1}{\left(1+\frac{r_0}{4R}\right)^{2}}; \quad (74)$$

$$\left(H_{\underline{2}}^{\prime 2}\right)_{is} = \left(H_{\underline{2}}^{2}\right)_{S} = \frac{1}{r} = \frac{1}{R\left(1+\frac{r_0}{4R}\right)^{2}}; \quad (75)$$

$$\left(H_{\underline{3}}^{\prime 3}\right)_{is} = \left(H_{\underline{3}}^{3}\right)_{S} = \frac{1}{r\sin\theta} = \frac{1}{R\sin\theta\left(1+\dfrac{r_0}{4R}\right)^2}. \tag{76}$$

The transformed tetrads coincide with the tetrads (66) in the Schwinger gauge for the Schwarzschild metric in the isotropic coordinates. In accordance with (35), the Hamiltonian with the tetrads (73) – (76) will coincide with the Hamiltonian (68).

Note that when defining the tetrad $\left(H_{\underline{0}}^{\prime 0}\right)_{is}$ (73), to preserve the condition of positivity $\sqrt{f_S}$ (47), it should fulfill the condition $R > \dfrac{r_0}{4}$.

Thus, although Hilbert's causality condition $g_{00} > 0$ in the transformed metric is fulfilled for the range $R \in (0, \infty)$ (except a single point $R = \dfrac{r_0}{4}$) the constraints imposed on the domain of the wave functions in the Hamiltonian (45) with the basic Schwarzschild metric hold true for the domain of the transformed Hamiltonian (68). The domain of the wave functions of Dirac's equation with the Schwarzschild metric in the isotropic coordinates equals

$$R > \frac{r_0}{4}. \tag{77}$$

### 4.2 Schwarzschild solution in spherical harmonic coordinates

The coordinates are

$$(t, R, \theta, \varphi). \tag{78}$$

The coordinate transformation is

$$r = R + \frac{r_0}{2}; \quad dr = dR. \tag{79}$$

The square of interval is

$$ds^2 = \frac{\left(1-\dfrac{r_0}{2R}\right)}{\left(1+\dfrac{r_0}{2R}\right)} dt^2 - \frac{\left(1+\dfrac{r_0}{2R}\right)}{\left(1-\dfrac{r_0}{2R}\right)} dR^2 - \left(1+\dfrac{r_0}{2R}\right)^2 R^2\left(d\theta^2 + \sin^2\theta d\varphi^2\right). \tag{80}$$

The values $(-g)$, $g_G$ and $\eta$ equal

$$-g = \left(1+\frac{r_0}{2R}\right)^4 R^4 \sin^2\theta, \tag{81}$$

$$g_G = \left(1+\frac{r_0}{2R}\right)^4, \tag{82}$$

$$\eta = \left(g_G\right)^{1/4}\left(g^{00}\right)^{1/4} = \frac{\left(1+\frac{r_0}{2R}\right)^{5/4}}{\left(1-\frac{r_0}{2R}\right)^{1/4}}. \tag{83}$$

The non-zero tetrads $\tilde{H}^\mu_{\underline{\alpha}}$ in the Schwinger gauge equal

$$\tilde{H}^0_{\underline{0}} = \sqrt{\frac{1+\frac{r_0}{2R}}{1-\frac{r_0}{2R}}};\ \tilde{H}^1_{\underline{1}} = \sqrt{\frac{1-\frac{r_0}{2R}}{1+\frac{r_0}{2R}}};\ \tilde{H}^2_{\underline{2}} = \frac{1}{R\left(1+\frac{r_0}{2R}\right)};\ \tilde{H}^3_{\underline{3}} = \frac{1}{R\sin\theta\left(1+\frac{r_0}{2R}\right)}. \tag{84}$$

The matrices $\tilde{\gamma}^\alpha$ equal

$$\tilde{\gamma}^0 = \sqrt{\frac{1+\frac{r_0}{2R}}{1-\frac{r_0}{2R}}}\gamma^{\underline{0}};\ \tilde{\gamma}^1 = \sqrt{\frac{1-\frac{r_0}{2R}}{1+\frac{r_0}{2R}}}\gamma^{\underline{1}};\ \tilde{\gamma}^2 = \frac{1}{R\left(1+\frac{r_0}{2R}\right)}\gamma^{\underline{2}};\ \tilde{\gamma}^3 = \frac{1}{\left(1+\frac{r_0}{2R}\right)R\sin\theta}\gamma^{\underline{3}}. \tag{85}$$

The self-adjoint Hamiltonian in the $\eta$- representation with the tetrads (84) equals

$$H_\eta = \sqrt{\frac{1-\frac{r_0}{2R}}{1+\frac{r_0}{2R}}}m\gamma^{\underline{0}} - i\frac{1-\frac{r_0}{2R}}{1+\frac{r_0}{2R}}\gamma^{\underline{0}}\gamma^{\underline{1}}\left(\frac{\partial}{\partial R} + \frac{1}{R}\right) -$$
$$-i\sqrt{\frac{1-\frac{r_0}{2R}}{1+\frac{r_0}{2R}}}\frac{1}{1+\frac{r_0}{2R}}\left[\gamma^{\underline{0}}\gamma^{\underline{2}}\frac{1}{R}\left(\frac{\partial}{\partial\theta}+\frac{1}{2}\text{ctg}\,\theta\right)+\gamma^{\underline{0}}\gamma^{\underline{3}}\frac{1}{R\sin\theta}\frac{\partial}{\partial\varphi}\right] - \tag{86}$$
$$-\frac{i}{2}\gamma^{\underline{0}}\gamma^{\underline{1}}\frac{\partial}{\partial R}\left(\frac{1-\frac{r_0}{2R}}{1+\frac{r_0}{2R}}\right).$$

The current components (69) for the present case equal

$$j^0 = \psi^+_\eta \frac{1}{\left(1+\frac{r_0}{2R}\right)^2}\psi_\eta, \tag{87}$$

$$j^r = j^\theta = 0, \tag{88}$$

$$j^\varphi = \psi^+_\eta \frac{\left(1-\frac{r_0}{2R}\right)^{1/2}}{\left(1+\frac{r_0}{2R}\right)^{7/2}R\sin\theta}\gamma^{\underline{0}}\gamma^{\underline{3}}\psi_\eta. \tag{89}$$

As a result of the coordinate transformation (79), the tetrads (84) transform in accordance with (41).

$$\left(H_{\underline{0}}^{\prime 0}\right)_{gr} = \frac{\partial t}{\partial t}\left(H_{\underline{0}}^{0}\right)_{S} = \frac{1}{\sqrt{f_S}} = \sqrt{\frac{1+\frac{r_0}{2R}}{1-\frac{r_0}{2R}}}; \tag{90}$$

$$\left(H_{\underline{1}}^{\prime 1}\right)_{gr} = \frac{\partial R}{\partial r}\left(H_{\underline{1}}^{1}\right)_{S} = \sqrt{f_S} = \sqrt{\frac{1-\frac{r_0}{2R}}{1+\frac{r_0}{2R}}}; \tag{91}$$

$$\left(H_{\underline{2}}^{\prime 2}\right)_{gr} = \left(H_{\underline{2}}^{2}\right)_{S} = \frac{1}{r} = \frac{1}{R\left(1+\frac{r_0}{2R}\right)}; \tag{92}$$

$$\left(H_{\underline{3}}^{\prime 3}\right)_{gr} = \left(H_{\underline{3}}^{3}\right)_{S} = \frac{1}{r\sin\theta} = \frac{1}{R\sin\theta\left(1+\frac{r_0}{2R}\right)}. \tag{93}$$

The transformed tetrads (90) – (93) coincide with the tetrads (84) in the Schwinger gauge for the Schwarzschild metric in the spherical harmonic coordinates. In accordance with (35), the Hamiltonian with the tetrads (90) – (93) will coincide with the Hamiltonian (86).

Just as for the metric in the isotropic coordinates (see Subsection 4.1), when defining the tetrads $\left(H_{\underline{0}}^{\prime 0}\right)_{gr}$ (90) and $\left(H_{\underline{1}}^{\prime 1}\right)_{gr}$ (91), preserving the condition of reality $f_S$ (47) requires that the condition $R > \frac{r_0}{2}$ be fulfilled. This also follows from Hilbert's causality condition $g_{00} > 0$ for the metric of interest (80).

The treatment presented in Subsections 4.1, 4.2 shows that the self-adjoint Hamiltonians for the Schwarzschild metrics in the isotropic and harmonic coordinates (68), (86) are equivalent to the basic Hamiltonian (45), except for the change in the domain of the wave functions $R > \frac{r_0}{4}$ for the metric (68) and $R > \frac{r_0}{2}$ for the metric (86). These changes are due to the coordinate transformations (60), (79).

For all the three Hamiltonians (45), (68), (86), the hermiticity condition (28), (54) is fulfilled with a corresponding redefinition of "event horizons": $r_0$ - Schwarzschild metric (42); $\frac{r_0}{2}$ - Schwarzschild metric in the harmonic coordinates (80); $\frac{r_0}{4}$ - Schwarzschild metric in the isotropic coordinates (61), (62).

## 5. Eddington-Finkelstein and Painleve-Gullstrand metrics

### 5.1 Eddington-Finkelstein solution

The coordinates are

$$(T, r, \theta, \varphi). \tag{94}$$

The coordinate transformation is

$$dT = dt + \frac{r_0}{r} \frac{dr}{f_S}. \tag{95}$$

The square of interval is

$$ds^2 = f_S dT^2 - 2\frac{r_0}{r} dTdr - \left(1 + \frac{r_0}{r}\right) dr^2 - r^2 \left(d\theta^2 + \sin^2\theta d\varphi^2\right). \tag{96}$$

The values $(-g)$, $g_G$ and $\eta$ equal

$$-g = r^4 \sin^2\theta, \tag{97}$$

$$g_G = 1, \tag{98}$$

$$\eta = \left(1 + \frac{r_0}{r}\right)^{1/4}. \tag{99}$$

The non-zero tetrads $\tilde{H}^\mu_{\underline{\alpha}}$ in the Schwinger gauge equal

$$\tilde{H}^0_{\underline{0}} = \sqrt{1 + \frac{r_0}{r}}; \quad \tilde{H}^1_{\underline{0}} = -\frac{\frac{r_0}{r}}{\sqrt{1 + \frac{r_0}{r}}}; \quad \tilde{H}^1_{\underline{1}} = \frac{1}{\sqrt{1 + \frac{r_0}{r}}};$$

$$\tilde{H}^2_{\underline{2}} = \frac{1}{r}; \quad \tilde{H}^3_{\underline{3}} = \frac{1}{r\sin\theta}. \tag{100}$$

The matrices $\tilde{\gamma}^\alpha$ equal

$$\tilde{\gamma}^0 = \sqrt{1 + \frac{r_0}{r}} \gamma^0; \quad \tilde{\gamma}^1 = -\frac{\frac{r_0}{r}}{\sqrt{1 + \frac{r_0}{r}}} \gamma^0 + \frac{1}{\sqrt{1 + \frac{r_0}{r}}} \gamma^1; \quad \tilde{\gamma}^2 = \frac{1}{r} \gamma^2; \quad \tilde{\gamma}^3 = \frac{1}{r\sin\theta} \gamma^3. \tag{101}$$

The self-adjoint Hamiltonian in the $\eta$- representation with the tetrads (100) equals [13]

$$H_\eta = \frac{m}{\sqrt{1+\frac{r_0}{r}}} \gamma^0 - i\gamma^0\gamma^1 \frac{1}{1+\frac{r_0}{r}} \left( \frac{\partial}{\partial r} + \frac{1}{r} + \frac{r_0}{2r^2} \frac{1}{1+\frac{r_0}{r}} \right) -$$

$$-i\gamma^0\gamma^2 \frac{1}{\sqrt{1+\frac{r_0}{r}}} \frac{1}{r}\left(\frac{\partial}{\partial\theta} + \frac{1}{2}\text{ctg}\,\theta\right) - i\gamma^0\gamma^3 \frac{1}{\sqrt{1+\frac{r_0}{r}}} \frac{1}{r\sin\theta}\frac{\partial}{\partial\varphi} + \quad (102)$$

$$+i\frac{r_0}{r}\frac{1}{1+\frac{r_0}{r}}\left(\frac{\partial}{\partial r} + \frac{1}{r} - \frac{1}{2r\left(1+\frac{r_0}{r}\right)}\right).$$

The current components (69) equal

$$j^0 = \psi_\eta^+ \psi_\eta, \quad (103)$$

$$j^r = \psi_\eta^+ \left( -\frac{\frac{r_0}{r}}{1+\frac{r_0}{r}} + \frac{\gamma^0\gamma^1}{1+\frac{r_0}{r}} \right) \psi_\eta = 0, \quad (104)$$

$$j^\theta = 0, \quad (105)$$

$$j^\varphi = \psi_\eta^+ \frac{1}{\left(1+\frac{r_0}{r}\right)^{1/2} r\sin\theta} \psi_\eta. \quad (106)$$

We obtain the Hamiltonian (102) by direct transformation of the basic Hamiltonian (45) with the tetrads (43). In case of the coordinate transformation (95), the non-zero tetrads transformed in accordance with (41) equal

$$\left(H_{\underline{0}}^{\prime 0}\right)_{E-F} = \frac{\partial T}{\partial t}\left(H_{\underline{0}}^0\right)_S = \frac{1}{\sqrt{f_S}}; \quad (107)$$

$$\left(H_{\underline{1}}^{\prime 0}\right)_{E-F} = \frac{\partial T}{\partial r}\left(H_{\underline{1}}^1\right)_S = \frac{\frac{r_0}{r}}{\sqrt{f_S}}; \quad (108)$$

$$\left(H_{\underline{1}}^{\prime 1}\right)_{E-F} = \frac{\partial r}{\partial r}\left(H_{\underline{1}}^1\right)_S = \sqrt{f_S}; \quad (109)$$

$$\left(H_{\underline{2}}^{\prime 2}\right)_{E-F} = \left(H_{\underline{2}}^2\right)_S = \frac{1}{r}; \quad (110)$$

$$\left(H_{\underline{3}}^{\prime 3}\right)_{E-F} = \left(H_{\underline{3}}^{\prime 3}\right)_S = \frac{1}{r\sin\theta}. \quad (111)$$

As a result of the transformation (95), as compared with the tetrads (43), there appears an additional non-zero tetrad $\left(H_{\underline{1}}^{\prime 0}\right)_{E-F}$ (108).

Further, using the Lorentz transformations (15), (16), (21), (22), we reduce the tetrads (107) – (111) to the tetrads in the Schwinger gauge (100). The non-zero quantities $\Lambda_{\underline{\alpha}}^{\underline{\beta}}(r)$ in (15), (16) will be equal to

$$\Lambda_{\underline{0}}^{\underline{0}} = \Lambda_{\underline{1}}^{\underline{1}} = \frac{1}{\sqrt{f_S}\sqrt{1+\frac{r_0}{r}}}; \quad \Lambda_{\underline{0}}^{\underline{1}} = \Lambda_{\underline{1}}^{\underline{0}} = -\frac{\frac{r_0}{r}}{\sqrt{f_S}\sqrt{1+\frac{r_0}{r}}}. \tag{112}$$

As a result of this two-step transformation, the basic Hamiltonian (45) transforms into (102). The expressions, (107) – (111), (112) contain expression $\sqrt{f_S} = \sqrt{1-\frac{r_0}{r}}$, which is real in the basic metric (42) (see (47)). This implies that the domain of the wave functions of the Hamiltonian (102), as before, is

$$r \in (r_0, \infty). \tag{113}$$

Hilbert's causality condition $g_{00} > 0$ for the Eddington-Finkelstein metric (96) also results in the domain $r > r_0$.

The radial current component (104) is zero. So far as the Lorenz transformation preserves the values of Dirac's currents, the equality $j^r = 0$ (104) can be easily obtained in frame of reference with tetrad vectors (107) - (111) considering both (51) and constancy of the wave functions at coordinate transformations (95). Therefore, for the Eddington-Finkelstein metric, the hermiticity condition for the Hamiltonians (28), (54) is fulfilled.

### 5.2 Painleve-Gullstrand solution

The coordinates are

$$(T, r, \theta, \varphi). \tag{114}$$

The coordinate transformation is

$$dT = dt + \sqrt{\frac{r_0}{r}}\frac{1}{f_S}dr. \tag{115}$$

The square of interval is

$$ds^2 = f_S dT^2 - 2\sqrt{\frac{r_0}{r}}dTdr - dr^2 - r^2\left(d\theta^2 + \sin^2\theta d\varphi^2\right). \tag{116}$$

The values $(-g)$, $g_G$ and $\eta$ equal

$$-g = r^4 \sin^2\theta, \tag{117}$$

$$g_G = 1, \tag{118}$$

$$\eta = 1. \tag{119}$$

The non-zero tetrads $\tilde{H}^\mu_{\underline{\alpha}}$ in the Schwinger gauge equal

$$\tilde{H}^0_{\underline{0}} = 1; \quad \tilde{H}^1_{\underline{0}} = -\sqrt{\frac{r_0}{r}}; \quad \tilde{H}^1_{\underline{1}} = 1; \quad \tilde{H}^2_{\underline{2}} = \frac{1}{r}; \quad \tilde{H}^3_{\underline{3}} = \frac{1}{r\sin\theta}. \tag{120}$$

The matrices $\tilde{\gamma}^\alpha$ equal

$$\tilde{\gamma}^0 = \gamma^{\underline{0}}; \quad \tilde{\gamma}^1 = -\sqrt{\frac{r_0}{r}}\gamma^{\underline{0}} + \gamma^{\underline{1}}; \quad \tilde{\gamma}^2 = \frac{1}{r}\gamma^{\underline{2}}; \quad \tilde{\gamma}^3 = \frac{1}{r\sin\theta}\gamma^{\underline{3}}. \tag{121}$$

The self-adjoint Hamiltonian in the $\eta$- representation with the tetrads (120) equals [28], [13]

$$H_\eta = \gamma^{\underline{0}}m - i\gamma^{\underline{0}}\left\{\gamma^{\underline{1}}\left(\frac{\partial}{\partial r} + \frac{1}{r}\right) + \gamma^{\underline{2}}\frac{1}{r}\left(\frac{\partial}{\partial\theta} + \frac{1}{2}\text{ctg}\theta\right) + \right.$$
$$\left. +\gamma^{\underline{3}}\frac{1}{r\sin\theta}\frac{\partial}{\partial\varphi}\right\} + i\sqrt{\frac{r_0}{r}}\left(\frac{\partial}{\partial r} + \frac{3}{4}\frac{1}{r}\right). \tag{122}$$

The current components (69) equal

$$j^0 = \psi^+_\eta \psi_\eta, \tag{123}$$

$$j^r = \psi^+_\eta\left(-\sqrt{\frac{r_0}{r}} + \gamma^{\underline{0}}\gamma^{\underline{1}}\right)\psi_\eta = 0, \tag{124}$$

$$j^\theta = 0, \tag{125}$$

$$j^\varphi = \psi^+_\eta \frac{1}{r\sin\theta}\gamma^{\underline{0}}\gamma^{\underline{3}}\psi_\eta. \tag{126}$$

We obtain the Hamiltonian (122) by direct transformation of the basic Hamiltonian (45) with the tetrads (43). In case of the coordinate transformation (115), the non-zero tetrads transformed in accordance with (41) equal

$$\left(H'^0_{\underline{0}}\right)_{P-G} = \frac{\partial T}{\partial t}\left(H^0_{\underline{0}}\right)_S = \frac{1}{\sqrt{f_S}}, \tag{127}$$

$$\left(H'^0_{\underline{1}}\right)_{P-G} = \frac{\partial T}{\partial r}\left(H^1_{\underline{1}}\right)_S = \frac{\sqrt{\frac{r_0}{r}}}{\sqrt{f_S}}, \tag{128}$$

$$\left(H'^1_{\underline{1}}\right)_{P-G} = \frac{\partial r}{\partial r}\left(H^1_{\underline{1}}\right)_S = \sqrt{f_S}, \tag{129}$$

$$\left(H'^2_{\underline{2}}\right)_{P-G} = \left(H^2_{\underline{2}}\right)_S = \frac{1}{r}, \tag{130}$$

$$\left(H'^3_{\underline{3}}\right)_{P-G} = \left(H^3_{\underline{3}}\right)_S = \frac{1}{r\sin\theta}. \tag{131}$$

As a result of the transformation (115), as compared with the tetrads (43), there appears an additional non-zero tetrad $\left(H_{\underline{1}}^{\prime 0}\right)_{P-G}$ (128).

Further, using the Lorentz transformation (15), (16), (21), (22), we reduce the tetrads (127) – (131) to the tetrads in the Schwinger gauge (120). The non-zero quantities $\Lambda_{\underline{\alpha}}^{\underline{\beta}}(r)$ in (15), (16) will be equal to

$$\Lambda_{\underline{0}}^{\underline{0}} = \Lambda_{\underline{1}}^{\underline{1}} = \frac{1}{\sqrt{f_S}}; \quad \Lambda_{\underline{0}}^{\underline{1}} = \Lambda_{\underline{1}}^{\underline{0}} = -\frac{\sqrt{\frac{r_0}{r}}}{\sqrt{f_S}}. \tag{132}$$

As a result of this two-step transformation, the basic Hamiltonian (45) transforms into (122).

The expressions, (127) – (129), (132) contain expression $\sqrt{f_S} = \sqrt{1 - \frac{r_0}{r}}$, which is real in the basic metric (42) (see (47)). This implies that the domain of the wave functions of the Hamiltonian (122), as before, is

$$r \in (r_0, \infty). \tag{133}$$

Hilbert's causality condition $g_{00} > 0$ for the Painleve-Gullstrand metric (116) also results in the domain $r > r_0$.

The radial current component (124) is zero. So far as the Lorenz transformation preserves the values of Dirac's currents, this equality can be easily obtained in a system of tetrad vectors (127) - (131) considering both (51) and constancy of the wave functions at coordinate transformation (115). Therefore, for the Painleve-Gullstrand metric, the hermiticity condition for the Hamiltonians (28), (54) is fulfilled.

The analysis of Dirac's Hamiltonians in the Eddington-Finkelstein (96) and Painleve-Gullstrand (116) gravitational fields shows that their domains of the wave functions are the same as the domain of the wave functions of the basic Hamiltonian (45) in the Schwarzschild field

$$r \in (r_0, \infty).$$

This is due to the fulfillment of both Hilbert's condition $g_{00} > 0$ and reality condition $\sqrt{f_S} = \sqrt{1 - \frac{r_0}{r}}$ in the course of the direct two-step transformations of the basic Hamiltonian (45) to self-adjoint Hamiltonians in the $\eta$-representation for the Eddington-Finkelstein and Painleve-Gullstrand solutions.

Expanding the domain to

$$r \in (0, \infty),$$

as in [28], is improper.

For the Eddington-Finkelstein (96) and Painleve-Gullstrand (116) metrics the hermiticity of the Hamiltonians (102), (122) leads to possible existence of stationary bound states of spin-half particles.

### 6. Lemaitre-Finkelstein and Kruskal metrics

#### 6.1 Lemaitre-Finkelstein solution

The coordinates are

$$(T, R, \theta, \varphi). \tag{134}$$

The coordinate transformations are

$$dT = dt + \frac{\sqrt{\frac{r_0}{r}}}{f_S} dr, \quad dR = dt + \frac{dr}{f_S \sqrt{\frac{r_0}{r}}}, \tag{135}$$

$$R = T + \frac{2}{3} \frac{r^{3/2}}{r_0^{1/2}}, \quad r = \left[\frac{3}{2}(R-T)\right]^{2/3} r_0^{1/3}. \tag{136}$$

The square of interval is

$$ds^2 = dT^2 - \frac{dR^2}{\left[\frac{3}{2r_0}(R-T)\right]^{2/3}} - \left[\frac{3}{2}(R-T)\right]^{4/3} r_0^{2/3} \left(d\theta^2 + \sin^2\theta d\varphi^2\right). \tag{137}$$

The domain of $T, R$ in (137) is

$$R > T. \tag{138}$$

The values $(-g)$, $g_G$ and $\eta$ equal

$$-g = \left[\frac{3}{2}(R-T)\right]^2 r_0^2 \sin^2\theta, \tag{139}$$

$$g_G = \left[\frac{3}{2}(R-T)\right]^2 \frac{r_0^2}{R^4}, \tag{140}$$

$$\eta = (g_G)^{1/4} (g^{00})^{1/4} = \left(\left[\frac{3}{2}(R-T)\right]^2 \frac{r_0^2}{R^4}\right)^{1/4}. \tag{141}$$

The non-zero components of the tetrad vectors $\tilde{H}_{\underline{\alpha}}^{\mu}$ in the Schwinger gauge equal:

$$\tilde{H}^0_{\underline{0}} = 1;\ \tilde{H}^1_{\underline{1}} = \left[\frac{3}{2r_0}(R-T)\right]^{\frac{1}{3}};\ \tilde{H}^2_{\underline{2}} = \frac{1}{\left[\frac{3}{2}(R-T)\right]^{\frac{2}{3}} r_0^{\frac{1}{3}}};\ \tilde{H}^3_{\underline{3}} = \frac{1}{\left[\frac{3}{2}(R-T)\right]^{\frac{2}{3}} r_0^{\frac{1}{3}} \sin\theta}. \quad (142)$$

The matrices $\tilde{\gamma}^\alpha$ equal

$$\tilde{\gamma}^0 = \gamma^{\underline{0}};\ \tilde{\gamma}^1 = \left[\frac{3}{2r_0}(R-T)\right]^{\frac{1}{3}} \gamma^{\underline{1}};\ \tilde{\gamma}^2 = \frac{1}{\left[\frac{3}{2}(R-T)\right]^{\frac{2}{3}} r_0^{\frac{1}{3}}} \gamma^{\underline{2}};\ \tilde{\gamma}^3 = \frac{1}{\left[\frac{3}{2}(R-T)\right]^{\frac{2}{3}} r_0^{\frac{1}{3}} \sin\theta} \gamma^{\underline{3}}. \quad (143)$$

The self-adjoint Hamiltonian in the $\eta$ - representation with the tetrads (142) equals [13]

$$H_\eta = \gamma^{\underline{0}} m - i\gamma^{\underline{0}}\gamma^{\underline{1}} \left[\frac{3}{2r_0}(R-T)\right]^{\frac{1}{3}} \left(\frac{\partial}{\partial R} + \frac{1}{R}\right) - i\gamma^{\underline{0}}\gamma^{\underline{2}} \frac{1}{\left[\frac{3}{2}(R-T)\right]^{\frac{2}{3}} r_0^{\frac{1}{3}}} \times$$

$$\times \left(\frac{\partial}{\partial\theta} + \frac{1}{2}\mathrm{ctg}\,\theta\right) - i\gamma^{\underline{0}}\gamma^{\underline{3}} \frac{1}{\left[\frac{3}{2}(R-T)\right]^{\frac{2}{3}} r_0^{\frac{1}{3}} \sin\theta} \frac{\partial}{\partial\varphi} - \frac{i}{2} \gamma^{\underline{0}}\gamma^{\underline{1}} \frac{\partial}{\partial R}\left[\frac{3}{2}(R-T)\right]^{\frac{1}{3}}. \quad (144)$$

We obtain the Hamiltonian (144) by two-step transformation of the basic Hamiltonian (45) with the tetrads (43). As a result of the coordinate transformation (135), (136), the non-zero tetrads transformed in accordance with (41) equal

$$\left(H'^0_{\underline{0}}\right)_{L-F} = \frac{\partial T}{\partial t}\left(H^0_{\underline{0}}\right)_S = \frac{1}{\sqrt{f_S(R,T)}}, \quad (145)$$

$$\left(H'^1_{\underline{0}}\right)_{L-F} = \frac{\partial R}{\partial t}\left(H^0_{\underline{0}}\right)_S = \frac{1}{\sqrt{f_S(R,T)}}, \quad (146)$$

$$\left(H'^0_{\underline{1}}\right)_{L-F} = \frac{\partial T}{\partial r}\left(H^1_{\underline{1}}\right)_S = \frac{\sqrt{\frac{r_0}{r(R,T)}}}{\sqrt{f_S(R,T)}}, \quad (147)$$

$$\left(H'^1_{\underline{1}}\right)_{L-F} = \frac{\partial R}{\partial r}\left(H^1_{\underline{1}}\right)_S = \frac{1}{\sqrt{f_S(R,T)}\sqrt{\frac{r_0}{r(R,T)}}}, \quad (148)$$

$$\left(H'^2_{\underline{2}}\right)_{L-F} = \left(H^2_{\underline{2}}\right)_S = \frac{1}{r(R,T)}, \quad (149)$$

$$\left(H'^3_{\underline{3}}\right)_{L-F} = \left(H^3_{\underline{3}}\right)_S = \frac{1}{r(R,T)\sin\theta}. \quad (150)$$

In the variables $R, T$, in accordance with (136),

$$f_s(R,T) = 1 - \frac{r_0}{r(R,T)} = 1 - \left(\frac{r_0}{\frac{3}{2}(R-T)}\right)^{2/3}. \tag{151}$$

As compared with the tetrads (43), the transformations (135) give two additional non-zero tetrads (146), (147).

Further, using the Lorentz transformations (15), (16), (21), (22), we reduce the tetrads (145) – (150) to the tetrads in the Schwinger gauge (142). The non-zero quantities $\Lambda_{\underline{\alpha}}^{\beta}(R,T)$ in (15), (16) will be equal to

$$\Lambda_{\underline{0}}^{0} = \Lambda_{\underline{1}}^{1} = \frac{1}{\sqrt{f_s(R,T)}}; \quad \Lambda_{\underline{1}}^{0} = \Lambda_{\underline{0}}^{1} = -\sqrt{\frac{r_0}{r(R,T)}}\frac{1}{\sqrt{f_s(R,T)}}. \tag{152}$$

As a result of this two-step transformation, the basic Hamiltonian (45) transforms into (144).

Hilbert's causality condition $g_{00} > 0$ for the Lemaitre-Finkelstein metric (137) imposes no constrains on the domain of the wave functions of the Hamiltonian (144). However, the expressions (145) – (148), (152) contain expression $\sqrt{f_s(R,T)} = \sqrt{1 - \frac{r_0}{r(R,T)}}$, which is real and positive in the basic metric (42). This implies (see (151)) that in addition to (138) there exists an additional constraint on the domain of the wave functions of the Hamiltonian (144)

$$R - T > \frac{2}{3}r_0. \tag{153}$$

The Lemaitre-Finkelstein metric (167) is non-stationary and as distinct from the basic Hamiltonian (45), the Hamiltonian (144) in the $(R,T)$ variables is explicitly time-dependent, and it is impossible to define stationary bound states of Dirac particles in these variables.

**6.2 Kruskal solution**

The coordinates are

$$(v, u, \theta, \varphi). \tag{154}$$

The coordinate transformations are

$$\begin{aligned} u &= \sqrt{\frac{r_0}{r}}\sqrt{f_s}\exp\frac{r}{2r_0}\text{ch}\left(\frac{t}{2r_0}\right), \\ v &= \sqrt{\frac{r_0}{r}}\sqrt{f_s}\exp\frac{r}{2r_0}\text{sh}\left(\frac{t}{2r_0}\right), \end{aligned} \tag{155}$$

$$\frac{r_0}{r}\sqrt{f_s}\exp\frac{r}{2r_0} = u^2 - v^2,$$

$$\frac{t}{2r_0}\text{arcth}\frac{v}{u} = \frac{1}{2}\text{arcth}\frac{2uv}{u^2 + v^2},$$

(156)

$$du = \frac{1}{2r_0}\sqrt{\frac{r_0}{r}}\sqrt{f_s}\exp\frac{r}{2r_0}\text{sh}\left(\frac{t}{2r_0}\right)dt +$$

$$+ \frac{1}{2r_0}\frac{\sqrt{\frac{r_0}{r}}}{\sqrt{f_s}}\exp\frac{r}{2r_0}\text{ch}\left(\frac{t}{2r_0}\right)dr,$$

$$dv = \frac{1}{2r_0}\sqrt{\frac{r_0}{r}}\sqrt{f_s}\exp\frac{r}{2r_0}\text{ch}\left(\frac{t}{2r_0}\right)dt +$$

$$+ \frac{1}{2r_0}\frac{\sqrt{\frac{r_0}{r}}}{\sqrt{f_s}}\exp\frac{r}{2r_0}\text{sh}\left(\frac{t}{2r_0}\right)dr.$$

(157)

The square of interval is

$$ds^2 = f^2 dv^2 - f^2 du^2 - (r(u,v))^2(d\theta^2 + \sin^2 d\varphi^2),$$

$$(f(u,v))^2 = \frac{4r_0^3}{r(u,v)}\exp\left(-\frac{r(u,v)}{r_0}\right) = \text{function of } (u^2 - v^2).$$

(158)

The values $(-g)$, $g_G$ and $\eta$ equal

$$-g = (f(u,v))^4 (r(u,v))^4 \sin^2\theta,$$

(159)

$$g_G = \frac{(f(u,v))^4 (r(u,v))^4}{u^4},$$

(160)

$$\eta = (g_G \cdot g^{00})^{1/4} = (f(u,v))^{1/2}\frac{r(u,v)}{u}.$$

(161)

The non-zero components of the tetrad vectors $\tilde{H}_{\underline{\alpha}}^{\mu}$ in the Schwinger gauge equal:

$$\tilde{H}_{\underline{0}}^0 = \frac{1}{f(u,v)}; \quad \tilde{H}_{\underline{1}}^1 = \frac{1}{f(u,v)}; \quad \tilde{H}_{\underline{2}}^2 = \frac{1}{r(u,v)}; \quad \tilde{H}_{\underline{3}}^3 = \frac{1}{r(u,v)\sin\theta}.$$

(162)

The matrices $\tilde{\gamma}^\alpha$ equal

$$\tilde{\gamma}^0 = \frac{1}{f}\gamma^{\underline{0}}; \quad \tilde{\gamma}^1 = \frac{1}{f}\gamma^{\underline{1}}; \quad \tilde{\gamma}^2 = \frac{1}{r(u,v)}\gamma^{\underline{2}}; \quad \tilde{\gamma}^3 = \frac{1}{r(u,v)\sin\theta}\gamma^{\underline{3}}.$$

(163)

The self-adjoint Hamiltonian in the $\eta$- representation with the tetrads (162) equals

$$H_\eta = \gamma^0 f(u,v) m - i\gamma^0\gamma^1 \left(\frac{\partial}{\partial u} + \frac{1}{u}\right) - i\gamma^0\gamma^2 \frac{f(u,v)}{r(u,v)}\left(\frac{\partial}{\partial \theta} + \frac{1}{2}\text{ctg}\,\theta\right) - $$
$$-i\gamma^0\gamma^3 \frac{f(u,v)}{r(u,v)\sin\theta}\frac{\partial}{\partial \varphi}. \tag{164}$$

We obtain the Hamiltonian (164) by two-step transformation of the basic Hamiltonian (45) with the tetrads (43). As a result of the coordinate transformation (155) – (157), the non-zero tetrads transformed in accordance with (41) equal

$$\left(H_{\underline{0}}^{\prime 0}\right)_K = \frac{\partial v}{\partial t}\left(H_{\underline{0}}^0\right)_S = \text{ch}\left(\frac{t(u,v)}{2r_0}\right)\frac{1}{2r_0}\sqrt{\frac{r(u,v)}{r_0}}\exp\frac{r(u,v)}{2r_0}, \tag{165}$$

$$\left(H_{\underline{0}}^{\prime 1}\right)_K = \frac{\partial u}{\partial t}\left(H_{\underline{0}}^0\right)_S = \text{sh}\left(\frac{t(u,v)}{2r_0}\right)\frac{1}{2r_0}\sqrt{\frac{r(u,v)}{r_0}}\exp\frac{r(u,v)}{2r_0}, \tag{166}$$

$$\left(H_{\underline{1}}^{\prime 0}\right)_K = \frac{\partial v}{\partial r}\left(H_{\underline{1}}^1\right)_S = \text{sh}\left(\frac{t(u,v)}{2r_0}\right)\frac{1}{2r_0}\sqrt{\frac{r(u,v)}{r_0}}\exp\frac{r(u,v)}{2r_0}, \tag{167}$$

$$\left(H_{\underline{1}}^{\prime 1}\right)_K = \frac{\partial u}{\partial r}\left(H_{\underline{1}}^1\right)_S = \text{ch}\left(\frac{t(u,v)}{2r_0}\right)\frac{1}{2r_0}\sqrt{\frac{r(u,v)}{r_0}}\exp\frac{r(u,v)}{2r_0} \tag{168}$$

$$\left(H_{\underline{2}}^{\prime 2}\right)_K = \left(H_{\underline{2}}^2\right)_S = \frac{1}{r(u,v)}, \tag{169}$$

$$\left(H_{\underline{3}}^{\prime 3}\right)_K = \left(H_{\underline{3}}^3\right)_S = \frac{1}{r(u,v)\sin\theta}. \tag{170}$$

As compared with the tetrads (43), the transformations (155) - (157) give two additional non-zero tetrads (166), (167).

Further, using the Lorenz transformation (15), (16), (21), (22), we reduce the tetrads (165) – (170) to the tetrads in the Schwinger gauge (162). The non-zero quantities $\Lambda_{\underline{\alpha}}^{\underline{\beta}}(u,v)$ in (15), (16) will be equal to

$$\Lambda_{\underline{0}}^{\underline{0}} = \Lambda_{\underline{1}}^{\underline{1}} = \frac{1}{f}\text{ch}\left(\frac{t(u,v)}{2r_0}\right)\frac{1}{2r_0}\sqrt{\frac{r(u,v)}{r_0}}\exp\frac{r(u,v)}{2r_0};$$
$$\Lambda_{\underline{0}}^{\underline{1}} = \Lambda_{\underline{1}}^{\underline{0}} = -\frac{1}{f}\text{sh}\left(\frac{t(u,v)}{2r_0}\right)\frac{1}{2r_0}\sqrt{\frac{r(u,v)}{r_0}}\exp\frac{r(u,v)}{2r_0}. \tag{171}$$

As a result of this two-step transformation, the basic Hamiltonian (45) transforms into (164).

Hilbert's causality condition $g_{00} > 0$ for the Kruskal metric (158) imposes no constrains on the domain of the wave functions of the Hamiltonian (164). Similarly, there are no constraints in deriving the tetrads (165) – (170) in the first step of the transformation and at the Lorentz transformation in the second step of the transformation of the basic Hamiltonian (45). However, the domain of the wave functions of the transformed Hamiltonian is constrained when new variables $(u,v)$ are introduced. The equalities (155), (156) contain expression $f_S = 1 - \dfrac{r_0}{r(u,v)}$, which is real and positive in the basic metric (42) (see (46), (47)). This implies (see (155), (156)), that the following conditions should be fulfilled for the domain in the $(u,v)$ coordinates:

$$u^2 > v^2, \; u^2 \neq v^2 \neq 0. \tag{172}$$

In the $(u,v)$ plane, the domain of the wave functions of the Hamiltonian (164) is the right quadrant $u > |v|$. The lines $u = \pm v$ and the point $u = v = 0$ do not belong to the sought domain.

The Kruskal metric (158) is non-stationary and as distinct from the basic Hamiltonian (45), the Hamiltonian (164) in the $(u,v)$ variables is explicitly dependent on the temporal coordinate $v$, and it is impossible to define stationary bound states of Dirac particles in these variables.

If, for the particle motion under the "event horizon" $r < r_0$ by analogy with the metric (55), in the definition of the Kruskal variables (155), (156) we interchange the temporal and the radial coordinates, we will obtain purely imaginary coordinates $(u,v)$, which, given the real coordinates $\theta, \varphi$, is inacceptable for the quantum-mechanical treatment of the evolution of the "Dirac particle in external gravitational field" system.

The analysis of Dirac's Hamiltonians (144), (164) in the Lemaitre-Finkelstein (137) and Kruskal (158) gravitational fields shows that the domains of the wave functions are restricted, just as the original domain of the Hamiltonian (45) in the Schwarzschild field $(r \in (r_0, \infty))$.

In the Lemaitre-Finkelstein coordinates $(R,T)$ the constraint resolves into condition (153)

$$R - T > \frac{2}{3} r_0.$$

In the Kruskal coordinates $(u,v)$, the constraint equals

$$u > |v| > 0$$

In the Lemaitre-Finkelstein variables and in the Kruskal variables, the Hamiltonians (144), (164) explicitly depend on temporal coordinates. Therefore, for them, it is impossible to define stationary bound states of Dirac particles.

## 7. Conclusions

In the paper we analyze the quantum-mechanical equivalence of the metrics of a centrally symmetric uncharged gravitational field.

We considered the Schwarzschild metrics in the spherical [1], isotropic [2] and harmonic [3] coordinates; the Eddington-Finkelstein [8], [5] and the Painleve-Gullstrand [9], [10] metrics; and the Lemaitre-Finkelstein [4], [5] and Kruskal [6], [7] metrics. All the metrics were derived from the solution [1] by corresponding coordinate transformations.

For all the metrics we obtained self-adjoint Hamiltonians with a flat scalar product of the wave functions and tetrad vectors in the Schwinger gauge. In addition, the same Hamiltonians were obtained by direct two-step transformations of the basic Hamiltonian (45) for the Schwarzschild field in the $(t,r,\theta,\varphi)$ coordinates. First, in accordance with the coordinate transformations, for the metrics of interest, we transformed the basic Hamiltonian (45) with the tetrads (43). Further, if necessary, Lorentz transformations (15), (16), (21), (22) were performed to move to the tetrads in the Schwinger gauge.

For the metrics and the Hamiltonians of interest, the scope of the analysis included the domains of the wave functions of Dirac's equation, hermiticity of the Hamiltonians $((\Phi,H\Psi)=(H\Phi,\Psi))$, and the possibility of existence of stationary bound states of spin-half particles. As a result of the analysis, the following conclusions can be made:

1. For the basic Schwarzschild metric in the spherical coordinates $(t,r,\theta,\varphi)$, to fulfill Hilbert's causality condition $g_{00}>0$, the domain of the wave functions is restricted by the condition

$$r > r_0. \qquad (173)$$

For all the other metrics of interest, the condition (173) also manifests in new variables:

- Schwarzschild metric in the isotropic coordinates

$$R > \frac{r_0}{4}, \qquad (174)$$

- Schwarzschild metric in the harmonic coordinates

$$R > \frac{r_0}{2}, \qquad (175)$$

- Eddington-Finkelstein and Painleve-Gullstrand metrics

$$r > r_0, \qquad (176)$$

- Finkelstein-Lemaitre metric

$$R - T > \frac{2}{3} r_0, \qquad (177)$$

- Kruskal metric

$$u > |v| > 0. \qquad (178)$$

Inequalities (174) - (178) show that the "event horizon" $r_0$ in the basic Schwarzschild metric (42) manifests itself in new coordinates in all the metrics considered. The domains of the wave functions for all the metrics derived by coordinate transformations of the basic metric (42) subject to (173) don't include the singular point at the origin.

2. When considering the possibility of motion of Dirac particles under the "event horizon" $(r < r_0)$ for the Schwarzschild metric, due to the interchange between the temporal and radial coordinates [26], [27] (metric (55)), Dirac's Hamiltonian explicitly depends on time and is physically non-equivalent to the basic Hamiltonian (45) with the metric (42) and the domain (173). Hilbert's condition $g_{00} \neq 0$ does not allow cross-linking the wave functions at the "event horizon" $r = r_0$. The same replacement in the Kruskal variables (155), (156) leads to purely imaginary coordinates $u, v$ at real coordinates $\theta, \varphi$, that is physically inacceptable for the quantum-mechanical treatment of the "Dirac particle in gravitational field" system.

3. The self-adjoint Hamiltonians (68), (86) for the Schwarzschild metrics in the isotropic and harmonic coordinates are Hermitian, and for them, just as for the basic Hamiltonian (45), the existence of real stationary bound states of spin-half particles is possible.

4. The self-adjoint Hamiltonians (102), (122) for the Eddington-Finkelstein and Painleve-Gullstrand metrics are also Hermitian, and for them the existence of stationary bound states of spin-half particles is possible.

5. The self-adjoint Hamiltonians (144), (164) for the Lemaitre-Finkelsteinand Kruskal metrics have the explicit dependence on the temporal coordinates and stationary bound states of spin-half particles cannot be defined for these Hamiltonians.

The results of this study can be useful when addressing the issues related to the evolution of the universe and interaction of collapsars with surrounding matter.

**Acknowledgement**

We thank A.L. Novoselova for her significant technical help in the preparation of the paper.